\documentclass[showpacs,prd,amsmath,amssymb,preprint draft,nofootinbib]{revtex4-1}

\usepackage{bm}
\usepackage{float,color}
\usepackage{comment}
\usepackage{bbold}
\usepackage{dsfont}
\usepackage[pdftex]{graphicx}
\usepackage{physics}
\newcommand{\eq}[1]{\begin{align} #1 \end{align}}

\newcommand{\pa}{\partial}

\newcommand{\del}{\delta}

\usepackage{dsfont}

\begin{document}

	\title{Clock Time in Quantum Cosmology}
	
	\author{Marcello Rotondo}
	\email{marcello@gravity.phys.nagoya-u.ac.jp}
	\author{Yasusada Nambu}
	\email{nambu@gravity.phys.nagoya-u.ac.jp}
	\affiliation{Department of Physics, Graduate School of Science, Nagoya 
		University, Chikusa, Nagoya 464-8602, Japan}
	
	\date{ \today }

\begin{abstract}
	We consider the conditioning of the timeless solution to the Wheeler-DeWitt equation by a predefined matter clock state in the simple scenario of de Sitter universe. The resulting evolution of the geometrodynamical degree of freedom with respect to clock time is characterized by the ``Berry connection'' of the reduced geometrodynamical space, which relies on the coupling of the clock with the geometry. When the connection vanishes, the standard Schr\"odinger equation is obtained for the geometry with respect to clock time. When one considers environment-induced decoherence in the semi-classical limit, this condition is satisfied and clock time coincides with cosmic time. Explicit results for the conditioned wave functions for minimal clocks made up of two quantum harmonic oscillator eigen-states are shown.
\end{abstract}

\maketitle

\section{Introduction}

In the canonical approach to relativistic and non-relativistic quantum mechanics, where one supposes a
background classical geometry, classical time sticks out in
the Schr\"odinger (functional) equation as an external parameter used by the
observer. The apparent lack of external time
for the dynamics of canonical quantum gravity, described by the so-called ''problem of time'' put
forth by the Wheeler--DeWitt (WD) equation \cite{Dew67}
\begin{equation}
\widehat H \Psi = 0 \label{eq:hamQG} \, ,
\end{equation}
has therefore been one of the elements stimulating the research concerning the nature of time in an attempt toward understanding how the evolution of the quantum state of the universe $\Psi$ can be described when time and space themselves become dynamical variables \cite{Ban85,Zeh86,Unr89b,Hal96,And12}.

Different approaches to introducing an effective time variable to the canonical picture of gravity have been devised. In one class of such attempts, one tries to extract a physical variable $t$ to be used as an effective time variable and obtain a Schr\"odinger-like structure for the Hamiltonian $H$:
\begin{equation}
\widehat{H} \Psi = \left(\widehat{H}_t - i \pa_t \right) \Psi = 0 \label{eq:sch} \, .
\end{equation}

The structure $\widehat{H}_t + \widehat{P}_t$, with $\widehat{H}_t$ being a physical Hamiltonian describing the evolution with respect to the time variable $t$ and $\widehat{P}_t$ being the momentum canonically conjugated to time, is characteristic of time-reparameterized Hamiltonians in the classical theory, i.e., Hamiltonians where
coordinate time $t$ has been promoted to a dynamical variable
$t(\theta)$ that is dependent on some implicit, unobservable time $\theta$. The
absence of a structure like Equation \eqref{eq:sch} with respect to time in canonical gravity, already at the classical level, is due to the fact that the theory is built on space--time diffeomorphism invariance to begin with, and the attempts to identify a canonical momentum to use as a generator of time translations are plagued by difficulties of various kinds and degrees (see, e.g., \cite{Kuc92,Ish93,Kie00}).

A rather general idea to extracting dynamics from a seemingly stationary
system has been proposed in what is sometimes called the
Page-and-Wootters approach or the conditional probability
interpretation (CPI) \cite{Pag83,Woo84}. In this approach, ``time'' evolution is read, 
under the condition that the total state Hamiltonian
\eq{
	\widehat{H} = \widehat{H}_C \otimes 1_R + 1_C \otimes \widehat{H}_R \label{eq:HRC}
}
is constrained, in the quantum correlations between the two partitions of the total state, i.e., a clock state (C) with the physical state (R) (the ``rest of the universe'') entangled with it. Time evolution emerges from the measurement of an observable clock of some
kind, whose reading conditions the physical state.
Attention to this approach seems to have undergone a recent revival, especially after some of the most important
criticisms levied against it in the past have been addressed, together with an experimental illustration of the mechanism \cite{Dol08,Gam09,Gio15,Mar17,Mor14}. The actual application of the CPI to timelessness in canonical quantum gravity,
which originally motivated it, still seems to be quite lacking. One~example is \cite{Gam04}, which may have implications for the present work because it also presents a treatment of quantum decoherence, and recent contributions that take into account the gravitational coupling of the clock are provided in a series of works \cite{Bry18a,Bry18b,Bry18c}. In the latter references, the importance of accounting for the coupling of quantum clocks with gravity in the CPI is addressed. Soon after the first version of this work appeared, one of the authors made us aware of another recent contribution to the general case of coupled clocks in the context of non-relativistic quantum mechanics \cite{Smi17}. We redirect the reader to this and the previously cited works for a good introduction to the CPI and some aspects of the effects of coupling quantum clocks with gravity. Indeed, one important characteristic that is generally put forth as a requirement for a good clock is that it weakly interacts with the system whose evolution it describes. When we take gravity as partaking in the dynamics, though, it couples to all forms of energy, including the clock, and while the coupling might become weak for some chosen cases (e.g., when the clock has conformal coupling to a conformally invariant geometry) or approximations, it may generally be strong or even dominant in the quantum regime. In this regard, note that the results of the mentioned study \cite{Smi17} do not apply straightforwardly to the case of canonical quantum gravity considered here in that it is not sufficient to just include a contribution to the total Hamiltonian coming from the coupling: the completely free Hamiltonian contribution from the clock must also be explicitly excluded. Concerning the importance of the coupling, note also that when we consider how classical time emerges from the timeless WD Equation \eqref{eq:hamQG} in the semiclassical limit, we see that it comes to be defined as the parameter along the classical trajectories of the gravitational degrees of freedom in the WKB approximation. The origin of the classical time of the (functional) Schr\"odinger equation therefore lies precisely in the coupling between geometry and matter fields. 

In the present work, rather than trying to recover an equation of the Schr\"odinger type (Equation~\eqref{eq:sch}) or a notion of time through dynamical observables, we study how, starting from a definition of the clock time, the resulting conditioned state for the geometry evolves with respect to it in the general case where coupling cannot be neglected.
We treat the simple example of the FLRW mini-superspace model with a homogeneous massive scalar field minimally coupled to gravity. 
In Section \ref{sec:model}, we introduce the model and expand the solution of the WD equation in the eigenstates of the matter Hamiltonian. In Section \ref{sec:emergence}, we discuss how time emerges in the semiclassical limit of gravity and how it appears for conditioning by a predefined matter clock in the quantum regime. In Section \ref{sec:solution}, we show the explicit solution for minimal working clocks. Final observations and conclusions are provided~in~Section~\ref{sec:end}.
\section{Mini-Superspace Model}
\label{sec:model}
We consider the simple scenario of the mini-superspace for a
spatially flat FLRW universe of scale factor $a$,
\eq{
	ds^2 = -N(t)^2 dt^2 + a(t)^2 \delta_{ij} d x^i dx^j 
}
with a minimally coupled scalar field $\phi$ and a cosmological
constant $\Lambda$. The
classical action is \footnote{$$ \hbar = 1 \, , \quad \kappa^2=8\pi G= \frac{1}{M_p^2} \, .$$}
\begin{align}
	S &= \frac{1}{2\kappa^2} \int d^4x \sqrt{-g} \left( R - 2
	\Lambda \right) + \int d^4x \sqrt{-g} \left( -
	\frac{1}{2}\left(\pa_\mu\phi\right)^2 - U(\phi) \right) \, .
\end{align}

The Lagrangian is
\begin{align}
	L&=\frac{V}{\kappa^2}\left(-\frac{3a\dot
		a^2}{N}-Na^3\Lambda\right)
	+Va^3\left(\frac{\dot\phi^2}{2N}-NU\right)\\
	&=\frac{1}{\kappa^2}\left(-\frac{\dot\rho^2}{3N\rho}-N\rho\Lambda\right)+\rho
	\left(\frac{\dot\phi^2}{2N}-NU\right),
\end{align}
where $\, \dot{} \, := d/d t$, and $V$ denotes the comoving volume of the universe. We also introduce the physical volume of the universe $\rho:=Va^3$ as a
dynamical variable. The canonical momenta are
\begin{equation}
\pi_\rho=\frac{\pa
	L}{\pa\dot\rho}=-\frac{1}{\kappa^2}\frac{2\dot\rho}{3N\rho},
\quad\pi_\phi=\frac{\pa L}{\pa\dot\phi}=\frac{\rho\,\dot\phi}{N}.
\end{equation}

The Hamiltonian reads 
\begin{align}
	H & = \pi_\rho \dot{\rho} + \pi_\phi \dot{\phi} -
	L=N\,(H_\rho+H_\phi) \, ,
\end{align}
where 
\begin{equation}
H_\rho:=\rho\left(-\frac{3\kappa^2}{4}\pi_\rho^2+\frac{\Lambda}{\kappa^2}\right)
,\quad
H_\phi:= \frac{\pi_\phi^2}{2\rho} + \rho\, U(\phi).
\end{equation}

The metric is
\begin{equation}
ds^2=-N^2dt^2+(\rho/\rho_0)^{2/3}\,\delta_{ij}dx^idx^j,\quad \rho_0:=V \, . \label{eq:metric}
\end{equation}

The first-class Hamiltonian constraint
\begin{equation}
\delta H / \delta N=H_\rho+H_\phi= 0  \label{eq:WDWE1}
\end{equation}
becomes
\begin{equation}
\frac{1}{3}\left(\frac{\dot\rho}{N\rho}\right)^2=
\Lambda+\kappa^2\left[\frac{\rho^2}{2}
\left(\frac{\dot\phi}{N}\right)^2+U(\phi)\right] \, ,
\label{eq:FRW}
\end{equation}
which corresponds to the Friedmann equation for the  universe with a minimal
scalar field if we choose~$N=1$.

Constraint quantization of Equation \eqref{eq:WDWE1} for a physical state
$\Psi$ gives the WD equation,
\begin{equation}
(\widehat H_\rho+\widehat H_\phi) \Psi =0 \, .
\end{equation}

In the representation diagonalizing $(\rho, \phi)$, we have the WD
equation for the wave function $\Psi(\rho,\phi)$:
\begin{equation}
\left[\rho\left(\frac{\pa^2}{\pa\rho^2}+\lambda\right)
+\frac{4}{3\kappa^2}\left( -
\frac{1}{2\rho} \frac{\pa^2}{\pa\phi^2} +\rho\,U(\phi)\right)
\right] \Psi(\rho,\phi) 
= 0 \, , \label{eq:WD} 
\end{equation}
where $\lambda=4\Lambda/(3\kappa^4)$, and we have assumed the appropriate
ordering of the operator $\widehat\pi_\rho$ for simplicity~of~analysis.

For the free massive field $U(\phi)=\mu^2\phi^2/2$, the matter Hamiltonian is
\begin{equation}
\widehat H_\phi = -\frac{1}{2\rho}\frac{\pa^2}{\pa\phi^2}  + \frac{1}{2}\,
\rho\,\mu^2 \phi^2 \, . \label{eq:harm} 
\end{equation}

When we consider the eigenstates of the Hamiltonian 
$\widehat H_\phi$, the variable $\rho$ contained in $\widehat H_\phi$ can be
treated as an unknown external parameter, and the eigenvalue equation is
%
\begin{equation}
\widehat H_\phi \, \chi_n(\phi|\rho) = E_n \, \chi_n(\phi|\rho) \, ,
\end{equation}
where $\chi_n(\phi|\rho)$ are the energy eigenstates of $\widehat H_\phi$ for any value of $\rho$,
and the energy eigenstates can be determined by
\begin{equation}
\chi_n(\phi|\rho) = \frac{1}{\sqrt{2^n n!\sqrt{\pi}}} \left(
\mu\rho\right)^\frac{1}{4} e^{-\mu\rho\,
	\phi^2/2}\,  H_n\left(\sqrt{\mu\rho}\, \phi \right).
\label{eq:n_states} 
\end{equation}

$H_n(x)$ are the Hermite polynomials,
and the associated eigenvalues are 
\begin{equation}
E_n = \mu\left( n+\frac{1}{2}
\right), \quad n=0,1,2,\cdots \label{eq:E} \, . 
\end{equation}

The eigenmodes $\chi_n$ satisfy the
orthogonality and completeness relations
\begin{align}
	&(\chi_m,\chi_n):=\int_{-\infty}^{\infty}
	d\phi\,\chi_m^*(\phi|\rho)\,\chi_n(\phi|\rho)=\delta_{mn} \, , \label{eq:c1}\\
	&\sum_k\chi_k^*(\phi_1|\rho)\chi_k(\phi_2|\rho)=\del(\phi_1-\phi_2) \, . \label{eq:c2}
\end{align}

Then, we can generally expand the universal wave function
$\Psi(\rho,\phi)$ as
\begin{equation}
\Psi(\rho, \phi)=\sum_n\psi_n(\rho)\chi_n(\phi|\rho) \, .
\label{eq:Psi-a}
\end{equation}
%
where the components $\psi_n(\rho)$ of the expansion encode the information on the quantum state of the geometry for a given choice of the clock. These components are determined by substituting \eqref{eq:Psi-a} into~the~WD~Equation~\eqref{eq:WD}:
\begin{equation}
\sum_n\left[\rho\left(\pa_\rho^2\psi_n+\lambda\,\psi_n\right)
\chi_n+\frac{4}{3\kappa^2}\psi_n\,\widehat
H_\phi\chi_n\right]=-\sum_n\rho\left[2\pa_\rho\psi_n\,\pa_\rho\chi_n+
\psi_n\,\pa_\rho^2\chi_n\right] \, . \label{eq:WD1}
\end{equation}

After taking the inner product with $\chi_m$, the wave function of the
universe $\psi_n$ obeys
\begin{align}
	&\sum_{n}\left[\rho\sum_k\widehat{\mathcal{D}}_{mk}\widehat{\mathcal{D}}_{kn}
	+\left(\rho\lambda
	+\frac{4E_m}{3\kappa^2}\right)\del_{mn} - (\pa_\rho \chi_m, \pa_\rho \chi_n) + \sum_k \left( \pa_\rho \chi_m , \chi_k \right)\left( \chi_k , \pa_\rho \chi_n \right)\right]\psi_n=0 \, , \label{eq:WD2b}
\end{align}
where the covariant derivative is introduced as
\begin{equation}
\widehat{\mathcal{D}}_{mn}:=\del_{mn}\pa_\rho-iA_{mn} \, ,
\end{equation}
with the ``Berry'' connection \cite{Ber84,Ber85}
\begin{equation}
A_{mn}:=i(\chi_m,\pa_\rho\chi_n) \, . \label{eq:Amn}
\end{equation}

If we take into account the completeness of the energy eigenbasis $\{\chi_n\}$, the last two terms of Equation \eqref{eq:WD2b} vanish, and we obtain
	\begin{align}
	&\sum_{n}\left[\rho\sum_k\widehat{\mathcal{D}}_{mk}\widehat{\mathcal{D}}_{kn}
	+\left(\rho\lambda
	+\frac{4E_m}{3\kappa^2}\right)\del_{mn}\right]\psi_n=0 \, , \label{eq:WD2}
	\end{align}

As a result of the connection, different components $\psi_n$ of the expansion in \eqref{eq:Psi-a} become generally coupled to each other. For each component, the connection leads to a geometric phase, and the formal solution of Equation \eqref{eq:WD2} is given by
\footnote{
	For a function $\psi = B G$, $(\pa_\rho - i A)\psi = (\pa_\rho B - i A B) G + B \pa_\rho G = B \pa_\rho G$ for $\pa_\rho B = i A B$. The solution of this condition is $B = P \exp\left( i \int d\rho A(\rho)\right)$ .
}
\eq{
	\psi_n(\rho) = \sum_m \left[ P \exp \left(i \int d\rho A(\rho)\right)\right]_{nm} b_m G_m(\rho) \, , \label{eq:sol}
}
where $b_m$ represents constants, the symbol $P$ denotes a path ordering, and the functions $G_n$ satisfy
\eq{
	-\widehat{H}_\rho G_n = E_n G_n \, \label{eq:Gn}
}
which gives
\eq{
	G_n=\rho\,
	e^{-i\sqrt{\lambda}\,\rho}{}_1F_1[1+i\beta_n/\sqrt{\lambda},2,2i\sqrt{\lambda}\rho]
	\, ,\quad\beta_n=\frac{4\mu}{3\kappa^2}(n+1/2) \, ,\quad
	n=0\, , \, 1 \, , \, 2 \, , \, \dots 
}
with ${}_1F_1$ being Kummer's confluent hypergeometric function.

\section{Emergence of Time}
\label{sec:emergence}
\vspace{-6pt}
\subsection{WKB Time}

In the semiclassical limit, time naturally appears as a parameter along the superspace trajectories of the spatial geometry.
One considers the WKB ansatz for the wave function \cite{Kie94,Hal85}
\begin{equation}
\Psi(\rho, \phi) = \exp \left[ i \sum_{n=0}^\infty \left( \frac{3 \kappa^2}{2}\right)^{n-1} S_n(\rho,\phi)\right] \label{eq:WKB_PSI} \, . 
\end{equation}

By substituting in Equation \eqref{eq:WD} and equating each order of $\left(3 \kappa^2/2\right)^{p}$, one obtains
\eq{
	[p=-2]:& \quad \pa_\phi S_0 = 0 \, , \label{eq:p2} \\
	[p=-1]:& \quad \left( \pa_\rho S_0 \right)^2 = 3 \Lambda \, , \label{eq:p1} \\
	[\,\, p=0 \, \, \,]:& \quad \left( - \frac{1}{2\rho} \pdv[2]{}{\phi} + \rho U \right) e^{iS_1} = \rho \pa_\rho S_0 \pa_\rho S_1 \,  \label{eq:p0} \, , \\
	\dots \quad & \notag
}

Equations \eqref{eq:p2} and \eqref{eq:p1} give, respectively, $S_0 = S_0(\rho)$ and the Einstein--Hamilton--Jacobi equation for a de Sitter space, where matter contributions to the action are taken to be perturbative.
Then,~introducing WKB time as a parameter along the classical trajectories of $\rho(\tau)$,
\eq{
	\pa_\tau := - \rho (\pa_\rho S_0) \pa_\rho = - \rho \pi_\rho \pa_\rho \, , \label{eq:WKBT}
}
Equation \eqref{eq:p0} gives the functional Schr\"odinger equation
\eq{
	\mp i \pa_\tau \chi = \widehat{H}_\phi \chi  \,  \label{eq:STE}
}
for the matter wave functional $\chi := e^{iS_1}$ on the classical de Sitter background. Ambiguity in the sign of \eqref{eq:WKBT}, which determines the direction of the cosmological arrow of time, corresponds to the choice of sign in Equation \eqref{eq:p1}, where a positive sign gives the contracting de Sitter universe, and a negative sign yields the expanding one. Our definition of $G_n(\rho)$ (Equation \eqref{eq:Gn}) as a fundamental solution corresponds to a choice for fixing the sign ambiguity.
In the semiclassical approximation, the total wave function is
\eq{
	\Psi(\rho,\phi) \propto e^{\pm i \sqrt{\lambda} \rho} \chi(\rho,\phi) \, . 
}
%
\subsection{Scalar Field as a Clock}
Classical (WKB) time \eqref{eq:WKBT} relies on the coupling between geometry and matter and on the existence of trajectories for the geometrodynamical degrees of freedom. Therefore, it cannot be extended straightforwardly to the full quantum regime. In the present approach, rather than trying to extract a viable time variable from the dynamics of matter and geometry, we follow the original CPI proposal and start with the definition of the clock time through the scalar field and discuss the results of using the thus-defined time to track evolution. More specifically, given the Hamiltonian $\widehat H_\phi$, we define a ``clock time'' $T$ as a parameter of the following normalized clock wave function
\begin{equation}
\widetilde\chi(T,\phi|\rho)=\sum_{n} c_n\,e^{-i
	E_nT}\chi_n(\phi|\rho) \, ,\quad
\sum_n c_n^*c_n=1 \, , \label{eq:chi_t}
\end{equation}
where the sum is extended over non-vanishing coefficients $c_n\neq 0$ only. The state \eqref{eq:chi_t} formally satisfies the standard Schr\"odinger equation,
\begin{equation}
i \pa_T\widetilde\chi=\widehat H_\phi\,\widetilde\chi \, , \label{eq:sch_t}
\end{equation}
and the time variable $T$ in this sense is well-defined to begin with. Just as for quantum clocks in non-relativistic quantum mechanics, the time parameter $T$ can be estimated by measuring some physical observable of the clock $\widetilde\chi(T,\phi|\rho)$ and applying quantum estimation theory.
With respect to the parameter $T$, the quantum state of the geometry conditioned by the clock reading $T$ is then effectively described by
\eq{
	\widetilde{\psi}(\rho,T) :&= \left( \widetilde\chi(T,\phi|\rho) , \Psi(\rho,\phi, T)\right) = \sum_{m} c_m^*e^{i E_m T} \psi_m(\rho) =  \sum_{m,n} c_m^*b_n e^{i E_m T}  B_{mn} G_n(\rho) \, \label{eq:psim}
}
where for simplicity of notation we define the matrix of elements,
\eq{
	B_{mn}(\rho) := \left[ P \exp \left(i \int d\rho A(\rho)\right)\right]_{mn} \, . \label{eq:Bmn}
}

The exact form of the connection is
\begin{equation}
A_{mn}(\rho)=i(\chi_m,\pa_\rho\chi_n)=i
\frac{\alpha_{mn}}{4\rho} \, , \label{eq:A}
\end{equation}
where
\begin{align}
	\alpha_{mn}:&=4\rho\int_{-\infty}^{\infty}d\phi\,\chi_m^*(\phi|\rho)\,
	\pa_\rho\chi_n(\phi|\rho) \notag \\
	&=\sqrt{n(n-1)}\,\del_{m,n-2}-\sqrt{(n+2)(n+1)}\,\del_{m,n+2} \, .
	\label{eq:alpha}
\end{align}

Notice that for the present choice of Hamiltonian, the energy eigenvalues are independent of the parameter $\rho$ and are therefore not affected by the derivative in the connection.
The integral of the connection is
\begin{align}
	\int_{\rho_0}^\rho d\rho'
	A_{mn}(\rho')&= i \frac{\alpha_{mn}}{4} \,\ln(\rho/\rho_0) \, .
	\label{eq:int-connection}
\end{align}
where we can identify the arbitrary scale $\rho_0$ with the comoving volume of the universe introduced before. If we set this scale, for example, as the Planck scale, then $\rho < \rho_0$ belongs to the sub-Planckian, strong quantum regime.
\subsection{Evolution Equation}
The time evolution of state $\widetilde{\psi}(\rho,T) = \sum_n c_n^* e^{i E_n T} \psi_n$ is determined implicitly through its conditioning by the clock and is not generally of the Schr\"odinger type with the pure geometrodynamical Hamiltonian $H_\rho$. To derive the explicit dynamic law, one may start by noticing that
\eq{
	\frac{1}{\Delta T} \int_T^{T+\Delta T} dT' e^{i \mu n T'} = \delta_{n0} \, , \quad \Delta T := \frac{2 \pi}{\mu} \, ,
}
and, therefore,
\eq{
	& c_n^* \psi_n(\rho) = \frac{1}{\Delta T} \int_T^{T+\Delta T} dT' e^{-i E_n T'} \widetilde{\psi}(\rho,T') \, , \\
	& c_n^* E_n \psi_n(\rho) = -\frac{i}{\Delta T} \int_T^{T+\Delta T} dT' e^{-i E_n T'} \widetilde{\psi}(\rho,T') \pa_{T'} \widetilde{\psi}(\rho,T') \, .
}

This allows us to write
\eq{
	\widetilde{\psi}(\rho,T) & = \sum_n c_n^* e^{iE_nT}\psi_n \\
	& = \frac{1}{\Delta T} \sum_n \int_T^{T+\Delta T} dT' e^{-iE_n(T-T')} \widetilde{\psi}(\rho,T') \, .
}

A representation of the delta function in the interval $[T,T+\Delta T]$ is therefore
\eq{
	\delta(T'-T) := \frac{1}{\Delta T} \sum_n e^{-iE_n(T'-T)} \, .
}

Using the previous relations and assuming $c_n\neq0 \, \, \forall n$, the WD Equation \eqref{eq:WD2} gives the evolution equation of $\widetilde{\psi}$
\eq{
	i \pa_T \widetilde{\psi}(\rho,T) = \widehat{H}_\rho \widetilde{\psi}(\rho,T) + \frac{3\kappa^2}{4\Delta T} \rho \sum_{m\neq n} \frac{c_m^*}{c_n^*} \int_T^{T+\Delta T} dT' e^{i(E_m T - E_n T')} [\widehat{\mathcal{D}}^2]_{mn} \widetilde{\psi}(\rho,T') \label{eq:ev}\, .
}

Although the obtained equation has the structure of a differential-integral equation, when $[\mathcal{D}^2]_{mn}$ has only diagonal non-vanishing components (the connection is zero), the Schr\"odinger equation with Hamiltonian $\widehat{H}_\rho$ is recovered for $\widetilde{\psi}$.

The first term of the RHS of Equation \eqref{eq:ev} is inherited by the conditioned state independent of the coupling between the clock and the geometry and corresponds to the mechanism discussed by Page and Wootters \cite{Pag83,Woo84}. On the other hand, the second term depends on the coupling through the connection in the geometric phase term. Notice that the vanishing of this term does not necessarily require the coupling to be absent, and one may obtain for $\widetilde{\psi}$ the Schr\"odinger evolution generated by $\widehat{H}_\rho$ also when the coupling is strong. On the other hand, recovery of semiclassical time (WKB time) itself requires that a coupling between matter and geometry exists. 

In the semiclassical expansion of $\Psi(\rho,\phi)$ for solving the WD equation, the geometrodynamical variable $\rho$ and matter variable $\phi$ are treated, respectively, as ``heavy'' and ``fast'' degrees of freedom of the system. The evolution of the heavy degrees of freedom is described by the classical Einstein--Hamilton--Jacobi Equation \eqref{eq:p1} for the chosen metric, while the matter field describes quantum perturbations of the system. For a cyclic evolution of $\rho$, the various components of \eqref{eq:int-connection} become contributions to the so-called Berry phase acquired by the system, which has been discussed both in the context of non-relativistic quantum mechanics, where it was first introduced, as well as in quantum cosmology \cite{Bal90,Dat93,Kie07}. For discussions of the relation between the Berry connection and the emergence of semiclassical time, see also \cite{Bro89,Ber96,Ven90}. In the present case, though, the relevance of the phase ~\eqref{eq:int-connection} originates from the fact that it determines the coupling of different components of the expansion ~\eqref{eq:Psi-a} and the time evolution law of the geometrodynamical state.

When one takes into account the necessary coupling of gravity with ``environmental'' degrees of freedom \cite{Joo03}, the fast energy eigenmodes in the expansion \eqref{eq:Psi-a} decohere from each other, and one can neglect the off-diagonal elements of the connection \eqref{eq:Amn}, in which consists the so-called ``Born--Oppenheimer approximation''. Since the coherence between different components of the clock decays, the clock cannot be used any longer to track time because the quantum superposition is destroyed, and the time-dependent relative phases between distinct energy eigenstates are lost. This is reflected in the decoupling of the different indexes of $\psi_n$ in Equation \eqref{eq:WD2} and the diagonalization of the matrix \eqref{eq:Bmn}.
In this limit, we obtain an effective Schr\"odinger equation for each component of $\widetilde{\psi}$
\eq{
	\widehat{H}_\rho \psi_n = - E_n \psi_n
}
which is equivalent to the equation defining $G_n$ in the timeless picture (Equation \eqref{eq:Gn}). For a given {\itshape{n}}-branch with $E_n := E$, the solution is $\widetilde{\psi} \approx \exp \left[ i E T + i S(E,\rho)\right]$, and the classical trajectory is recovered by the condition
\eq{
	\text{const.} = \pa_E \left( E T + S \right) = T + \pa_E S(E,\rho) \, ,
}
from which the kinematic expression for $\rho = \rho(T)$ can be obtained. For the present case, it is possible to check that we obtain the classical equation for $\rho$ with a pure cosmological constant, with $N=1$ slice and clock time $T$ corresponding to cosmic time. It is important to observe that the correlation between $\rho$~and~$T$, as described by the resulting $\rho(T)$, emerges only in the classical limit and, due to its conditional nature with respect to the clock choice, it describes the classical ``rest of the universe'' alone and does not enter the definition of clock itself.

\section{Solution of the WD Equation with Clock}
\label{sec:solution}

We consider different clock models and show the behavior of the corresponding conditioned~state~$\widetilde{\psi}$.

\subsection{Clock with a Single Eigenstate}

To start, let us consider the total wave function with only the $m$th component,
\begin{equation}
\Psi(\rho,\phi)=\psi_m(\rho)\chi_m(\rho,\phi) \, .
\end{equation}

For this state, the connection vanishes $A = A_{mm} = 0$, and the wave function of the universe conditioned by the clock is simply
\begin{equation}
\widetilde \psi(T,\rho) = e^{i \mu (m + 1/2) T} G_m(\rho) \, .
\end{equation}

Thus, the wave function has a trivial time dependence given by an overall phase
factor, which is not measurable. The disappearance of clock time for the case of a single energy
eigenstate reflects the fact that a working clock needs a
superposition of at least two energy eigenstates to track the~time~evolution.

\subsection{Clock with Two Eigenstates }

Let us consider the minimal case of a working clock, made up of the only two eigenstates $0$~and~$1$ $(c_0=c_1=1/\sqrt{2})$. The timeless wave function will be
\eq{
	\Psi(\rho,\phi) = \frac{1}{\sqrt{2}} \left( \psi_0(\rho) \chi_0(\phi|\rho) + \psi_1(\rho) \chi_1(\phi|\rho) \right) \, .
}

Notice that when looking for the solutions $\psi_n$ (Equation \eqref{eq:sol}) satisfying Equation \eqref{eq:WD2}, we have assumed an expansion that includes all the eigenstates. This enables the use of the completeness relation to grant the vanishing of the extra $\rho^{-2}$-order term $ \sum_k (\pa_\rho \chi_m, \chi_k)(\chi_k, \pa_\rho\chi_n) -(\pa_\rho \chi_m, \pa_\rho \chi_n) $. We~may~consider the present case of a finite number of energy eigenstates as an approximation wherein all other coefficients $c_m$ are negligibly small and can be dropped from the equation. Also, in this case, the connection $A_{nm}$ ($n,m=0,1$) vanishes and $[e^{i\int d\rho A}]_{nm} = \mathds{1}_{nm}$. The conditional state becomes
\eq{
	\widetilde \psi(T,\rho) = \frac{e^{i\mu T /2}}{\sqrt{2}} \left( G_0(\rho) + e^{i \mu T} G_1(\rho)\right) \, , \label{eq:01}
}
where the relative phase makes the time dependence observable.

As discussed in the previous section, the simple time dependence of $\widetilde \psi(T,\rho)$ follows from the diagonality of the matrix \eqref{eq:Bmn}. For the choice of matter Hamiltonian \eqref{eq:harm}, diagonality is granted for $m \neq n \pm 2$. The simplest example in which this condition is not satisfied and $\widetilde \psi(T,\rho)$ takes a more complex time evolution is the case of eigenstates $n,m = {0,2}$:
\eq{
	\Psi = \frac{1}{\sqrt{2}} \left( \psi_0(\rho) \chi_0(\phi|\rho) + \psi_2(\rho) \chi_2(\phi|\rho) \right) \, .
}

The integral of the connection becomes
\begin{align}
	\int_{\rho_0}^\rho d\rho' A(\rho') &=\frac{i}{\sqrt{8}}\ln(\rho/\rho_0)
	\begin{pmatrix}
		0 &  1\\
		-1 & 0
	\end{pmatrix},
\end{align}
and
\begin{equation}
B =
\begin{pmatrix}
\cos\left(\frac{\ln(\rho/\rho_0)}{\sqrt{8}}\right) &
-\sin\left(\frac{\ln(\rho/\rho_0)}{\sqrt{8}}\right) \\
\sin\left(\frac{\ln(\rho/\rho_0)}{\sqrt{8}}\right) &
\cos\left(\frac{\ln(\rho/\rho_0)}{\sqrt{8}}\right) 
\end{pmatrix} \, , \label{eq:B}
\end{equation}
which is independent of the path ordering since the dependence on $\rho$ appears in the connection integral as an overall multiplicative factor.
The wave function of the universe conditioned by the clock (taking $b_n = 1/ \sqrt{2}$) becomes
\begin{align}
	&
	\widetilde\psi(T,\rho)
	= \frac{e^{i \mu T /2}}{2} \left[ \left( B_{00}(\rho) + e^{2i\mu T} B_{20}(\rho)\right)G_0(\rho) + \left( B_{02}(\rho) + e^{2i\mu T} B_{22}(\rho)\right)G_2(\rho) \right] \, . \label{eq:02}
	\end{align}

For the chosen clock Hamiltonian, the general time evolution of Equation \eqref{eq:ev} is simply
\eq{
	i \pa_T \widetilde{\psi}(\rho,T) = \widehat{H}_\rho \widetilde{\psi}(\rho,T) + \underbrace{\frac{3\kappa^2}{8} \left( \pa_\rho - \frac{1}{2\rho}\right) \sum_{m\neq n} c_m^* e^{iE_m T} \alpha_{mn} {\psi}_n(\rho)}_{X(\rho,T)} \, . \label{eq:ev2}
}

While the state \eqref{eq:01} exactly follows the Schr\"odinger equation with Hamiltonian $\widehat{H}_\rho$, for the time evolution of the state \eqref{eq:02}, the second term $X$ of Equation \eqref{eq:ev2} will generally not vanish.

\begin{figure}[H]
	\centering
	\includegraphics[width=0.49\linewidth,clip]{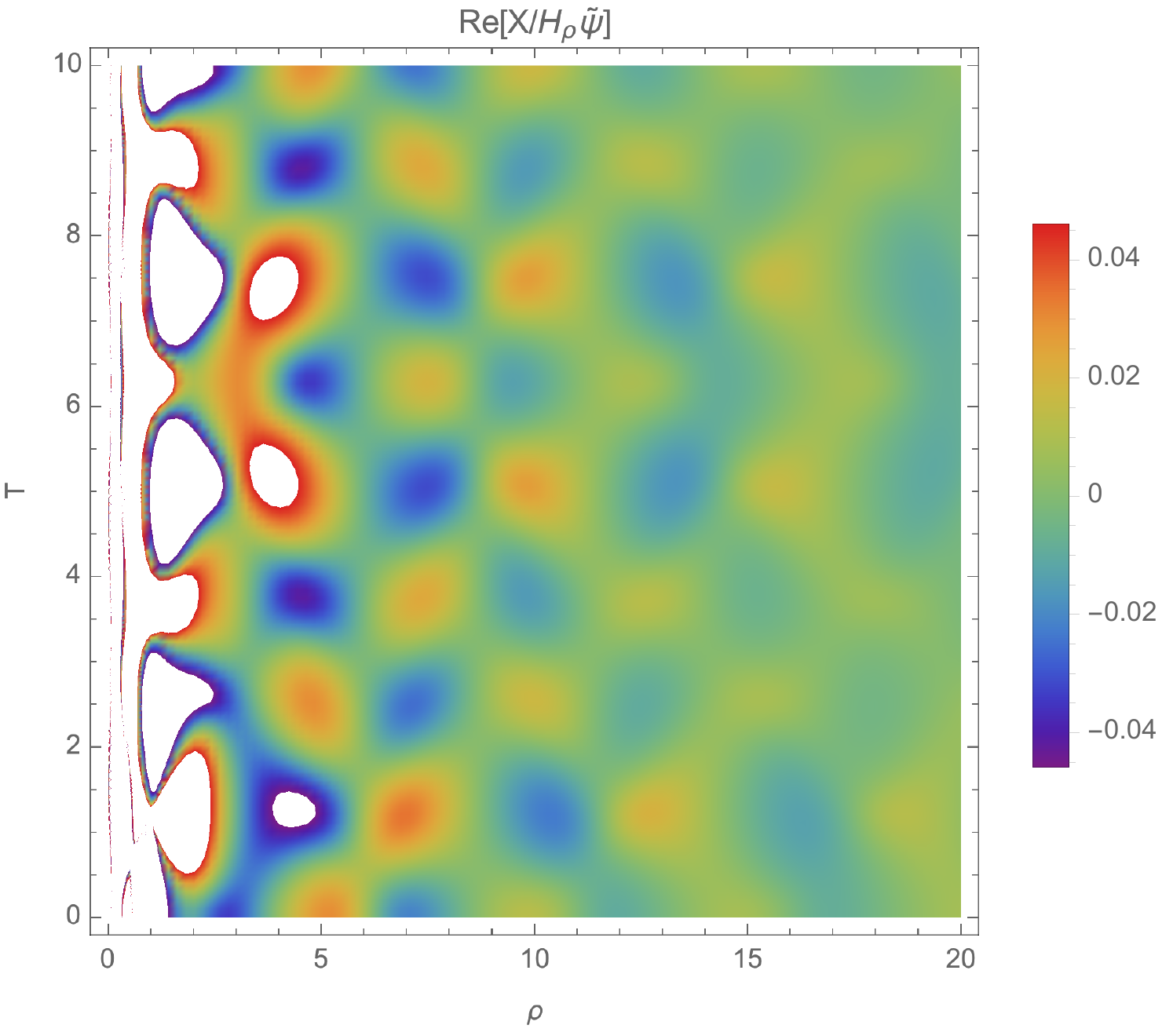}
	\includegraphics[width=0.49\linewidth,clip]{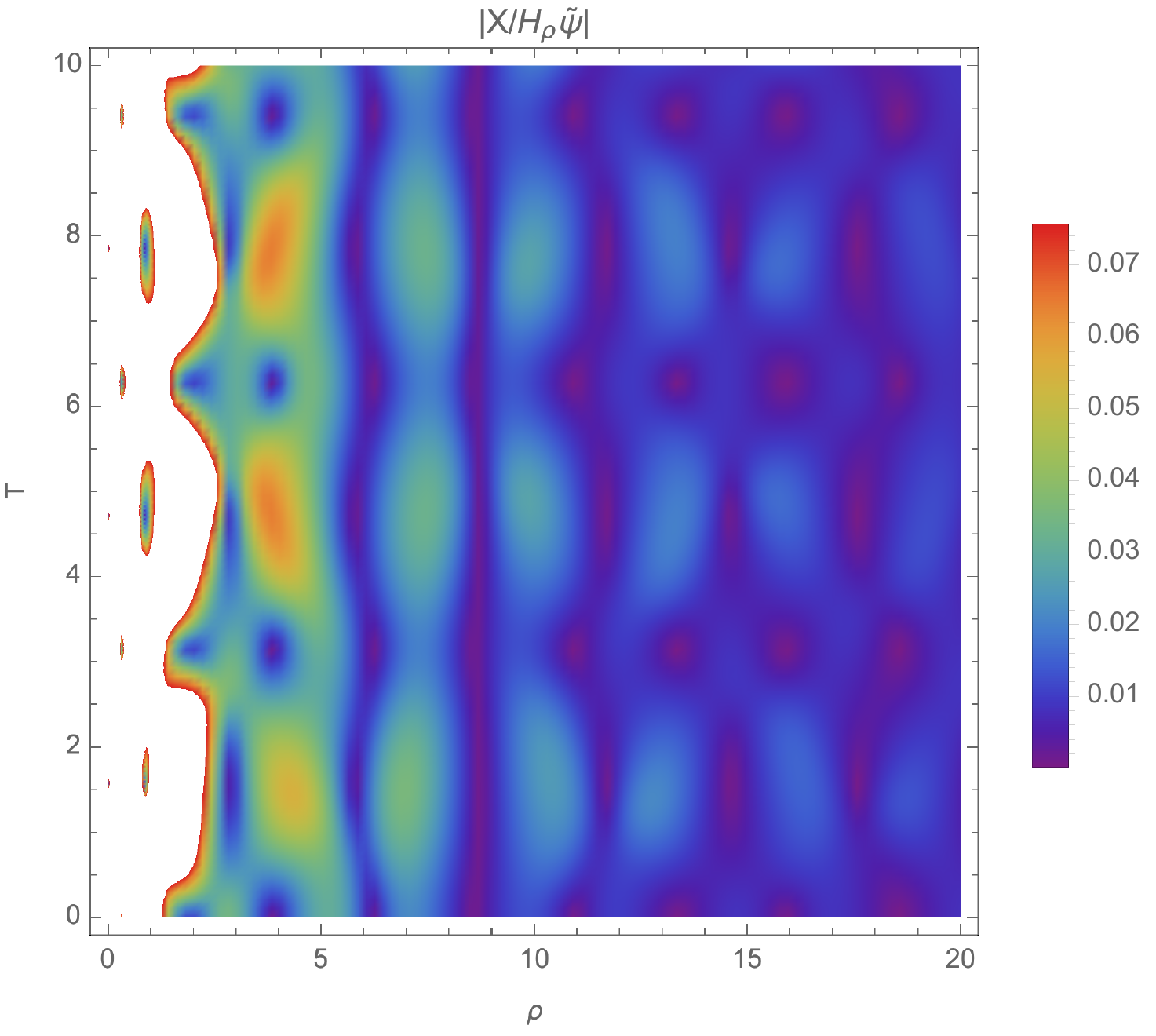}
	\caption{Plot of $\Re[X/H_\rho \widetilde \psi]$ (\textbf{left}) and $|X/H_\rho \widetilde \psi|$ (\textbf{right}) for unitary values of the physical constants $\lambda=\mu=\kappa=\rho_0=1 $ when the $\{0,2\}$-clock is employed.} \label{fig:fig1}
\end{figure}

The behavior of the ratio $X / H_\rho \widetilde \psi$ for the $\{0,2\}$-clock is shown in Figure \ref{fig:fig1}.
The values of $X(\rho,T)$ and $H_\rho \widetilde{\psi}$ are undetermined at $\rho = 0 \, \forall T$ because of the ``rotation'' $B(\ln(\rho/\rho_0)/\sqrt{8})$, whose argument diverges at $\rho = 0$. In this regime, Equation \eqref{eq:ev2} diverges from the Schr\"odinger evolution. On~the~other~hand,~for
\eq{
	\rho \gg \lambda^{-\frac{1}{2}} = \ell_P^2 \ell_H \, \label{eq:lim}
}
where $\ell_H^{-1} := 2 \sqrt{\Lambda / 3} $ is the de Sitter horizon scale and $\ell_P = \kappa$ is the Planck length, the ratio becomes negligibly small. In this limit, $G_m(\rho)$ can be approximated as
\eq{
	G_m(\rho) \approx g_m e^{i (\sqrt{\lambda}\rho + \frac{\beta_m}{\sqrt{\lambda}}\log (\sqrt{\lambda}\rho))} \, , \quad g_m = \frac{1}{4\sqrt{\lambda}}\frac{(2i)^{1+i \frac{\beta_m}{\sqrt{\lambda}}}}{\Gamma\left( 1 + i\frac{\beta_m}{\sqrt{\lambda}}\right)} \, .
}

On the other hand, $\pa_\rho B$ can be neglected in this limit and, neglecting the other contributions of order $\rho^{-1}$, $X(\rho,T)$ will be approximated as an oscillatory term of order $\kappa^{2}$:
\begin{equation}
X(\rho,T) \approx i \frac{3\kappa^2}{8} \sqrt{\frac{\lambda}{2}} \left( \left( B_{22} - e^{i2\mu T} B_{02} \right) e^{i\frac{\beta_2}{\sqrt{\lambda}} \log\left( \sqrt{\lambda}\rho\right)}g_2 + \left( B_{20} - e^{i2\mu T} B_{00} \right) e^{i\frac{\beta_0}{\sqrt{\lambda}} \log\left( \sqrt{\lambda}\rho\right)}g_0\right) e^{i(\frac{\mu}{2} T + \sqrt{\lambda}\rho )} \, ,
\end{equation}
which can be neglected compared with the growing contribution of the geometrodynamical Hamiltonian. Therefore, in the limit \eqref{eq:lim}, the Schr\"odinger equation is effectively recovered using this clock.
\section{Conclusions}
\label{sec:end}
In this article, we discuss the time dependence of the quantum state of the geometry obtained by conditioning the timeless solution of the Wheeler--DeWitt equation with a predefined clock state. The~resulting time dependence is generally not trivial, and the coupling between the clock and the geometry affects the quantum state through a geometric phase that couples different components of the expansion of the timeless state on the basis of the energy eigenstates of the clock. We derived an evolution law for the geometry with respect to the clock time that also holds when the coupling between the clock and the geometry cannot be neglected. A standard Schr\"odinger-type evolution generated by the geometrodynamical Hamiltonian is always recovered when the off-diagonal elements of the Berry connection, which determines the geometric phase, vanish. In the semiclassical limit when environment-induced decoherence is taken into account, the off-diagonal elements of the connection can effectively be neglected as well, and the different components of the clock decohere from each other. The disappearance of quantum superposition between different energy eigenstates of the clock makes it impossible to use it to track time: in this limit, observers will rely on the classical time parameter, which emerges in the WKB approximation of the total wave function. We show the explicit result for minimal working clocks made up of two harmonic oscillators. For a clock made by a superposition of the ground state and the first energy level (or, indeed, any two states that are not separated by two energy levels), the Schr\"odinger evolution is retrieved exactly. Using a superposition of the ground state and the second energy level harmonic oscillator, the evolution of the quantum wave function of the geometry presents a non-trivial deviation from the Schr\"odinger-type evolution generated by the pure geometrodynamical Hamiltonian. The Schr\"odinger equation is approximately recovered also for this clock for a physical volume much larger than the scale $ \ell_P^2 \ell_H $. This property is independent of the details of the decoherence mechanism.

Although we consider the simple case of an FLRW mini-superspace, the extension to other geometries and different matter Hamiltonians for the clock may proceed along the same lines.
We~focus on the effective time evolution resulting from the conditioning of the geometry by the clock state, but some important issues are not addressed and postponed to future work. For example, we do not consider the time evolution of the expectation values for $\tilde{\psi}$ (say, $\expval{\widehat \rho}{\tilde{\psi}}$), which would allow a discussion of the physical significance and applicability of the CPI for the study of early cosmology. Addressing this issue is conditioned to solving the problem of normalization for the wave function $\tilde{\psi}$ and recovering an interpretation for it as a probability amplitude density. Furthermore, we do not explicitly discuss the effect of the back-reaction of the clock on the state of the geometry. We expect that this would result in the classical limit in a modification of the Einstein--Hamilton--Jacobi equation, which may be derived from the effective Schr\"odinger evolution for $\rho$.

\vspace{12pt}

\acknowledgments{Y.N. was supported in part by JSPS KANENHI Grant No. 15K0573. M.R. gratefully acknowledges support from the Ministry of Education, Culture, Sports, Science and Technology (MEXT) of Japan. The authors would like to thank G. Venturi and A.R. Smith for suggesting useful references.}



\begin{thebibliography}{999}

\bibitem{Dew67}
DeWitt, B.S.
\newblock Quantum {theory} of {gravity}. {I}. {The} {canonical} {theory}.
\newblock {\em Phys. Rev.} {\bf 1967}, {\em 160},~1113--1148,
\newblock
  doi:10.1103/PhysRev.160.1113.

\bibitem{Ban85}
Banks, T.
\newblock {TCP}, quantum gravity, the cosmological constant and all that...
\newblock {\em Nucl. Phys. B} {\bf 1985}, {\em 249},~332--360,
\newblock
  doi:10.1016/0550-3213(85)90020-3.

\bibitem{Zeh86}
Zeh, H.D.
\newblock Emergence of classical time from a universal wavefunction.
\newblock {\em Phys. Lett. A} {\bf 1986}, {\em 116},~9--12,
\newblock
  doi:10.1016/0375-9601(86)90346-4.

\bibitem{Unr89b}
Unruh, W.G.; Wald, R.M.
\newblock Time and the interpretation of canonical quantum gravity.
\newblock {\em Phys. Rev. D} {\bf 1989}, {\em 40},~2598--2614,
\newblock
  doi:10.1103/PhysRevD.40.2598.

\bibitem{Hal96}
Halliwell, J.J.; Pérez-Mercader, J.; Zurek, W.H.
\newblock {\em Physical {Origins} of {Time} {Asymmetry}}; Cambridge~University~Press: Cambridge, UK, 1996.

\bibitem{And12}
Anderson, E.
\newblock Problem of time in quantum gravity.
\newblock {\em Annalen Phys.} {\bf 2012}, {\em 524},~757--786,
\newblock
  doi:10.1002/andp.201200147.

\bibitem{Kuc92}
Kuchař, K.V.
\newblock Time and {interpretations} of {quantum} {gravity}.
\newblock  \emph{Int. J. Mod. Phys. D} \textbf{2011}, \emph{20}, 3--86.

\bibitem{Ish93}
Isham, C.J.
\newblock Canonical {quantum} {gravity} and the {problem} of {time}. In {\em
  Integrable {Systems}, {Quantum} {Groups}, and {Quantum} {Field} {Theories}};
  Ibort, L.A., Rodríguez, M.A., Eds.; {NATO} {ASI} {Series}; Springer: Dordrecht, The~Netherlands, 1993; pp. 157--287,
\newblock
  doi:10.1007/978-94-011-1980-1\_6.

\bibitem{Kie00}
Kiefer, C.
\newblock Conceptual {issues} in {quantum} {cosmology}.
\newblock  In \emph{Towards {Quantum} {Gravity}}; Kowalski-Glikman, J., Ed.;  Lecture {notes} in {physics}; Springer: Berlin/Heidelberg, Germany, 2000; pp. 158--187.

\bibitem{Pag83}
Page, D.N.; Wootters, W.K.
\newblock Evolution without evolution: {Dynamics} described by stationary
  observables.
\newblock {\em Phys. Rev. D} {\bf 1983}, {\em 27},~2885--2892,
\newblock
  doi:10.1103/PhysRevD.27.2885.

\bibitem{Woo84}
Wootters, W.K.
\newblock ``{Time}'' replaced by quantum correlations.
\newblock {\em Int. J. Theor. Phys.} {\bf 1984}, {\em 23},~701--711,
\newblock
  doi:10.1007/BF02214098.

\bibitem{Dol08}
Dolby, C.E.
\newblock The conditional probability interpretation of the Hamiltonian
  Constraint. \emph{arXiv preprint} \textbf{2004}, gr-qc/0406034.

\bibitem{Gam09}
Gambini, R.; Porto, R.A.; Pullin, J.; Torterolo, S.
\newblock Conditional probabilities with Dirac observables and the problem of
  time in quantum gravity.
\newblock {\em Phys. Rev. D} {\bf 2009}, {\em 79},~041501, doi:10.1103/PhysRevD.79.041501.

\bibitem{Gio15}
Giovannetti, V.; Lloyd, S.; Maccone, L.
\newblock Quantum time.
\newblock {\em Phys. Rev. D} {\bf 2015}, {\em 92},~045033,
\newblock
  doi:10.1103/PhysRevD.92.045033.

\bibitem{Mar17}
Marletto, C.; Vedral, V.
\newblock Evolution without evolution and without ambiguities.
\newblock {\em Phys. Rev. D} {\bf 2017}, {\em 95},~043510,
\newblock
  doi:10.1103/PhysRevD.95.043510.

\bibitem{Mor14}
Moreva, E.; Brida, G.; Gramegna, M.; Giovannetti, V.; Maccone, L.; Genovese, M.
\newblock Time from quantum entanglement: {An} experimental illustration.
\newblock {\em Phys. Rev. A} {\bf 2014}, {\em 89},~052122,
\newblock
  doi:10.1103/PhysRevA.89.052122.

\bibitem{Gam04}
Gambini, R.; Porto, R.A.; Pullin, J.
\newblock A relational solution to the problem of time in quantum mechanics and
  quantum gravity: A fundamental mechanism for quantum decoherence.
\newblock {\em New J. Phys.} {\bf 2004}, {\em 6},~45,
\newblock
  doi:10.1088/1367-2630/6/1/045.

\bibitem{Bry18a}
Bryan, K.L.H.; Medved, A.J.M.
\newblock Realistic clocks for a {Universe} without time.
\newblock {\em Found. Phys.} {\bf 2018}, {\em 48},~48--59, doi:10.1007/s10701-017-0128-x,
\newblock arXiv:1706.02531.
 

\bibitem{Bry18b}
Bryan, K.L.H.; Medved, A.J.M.
\newblock Requiem for an ideal clock.
\newblock {\em arXiv:1803.02045} {\bf 2018},
\newblock arXiv:1803.02045.

\bibitem{Bry18c}
Bryan, K.L.H.; Medved, A.J.M.
\newblock The problem with ``{The} {problem} of {Time}''.
\newblock {\em arXiv:1811.09660} {\bf 2018}, arXiv:1811.09660.

\bibitem{Smi17}
Smith, A.R.H.; Ahmadi, M.
\newblock Quantizing time: {Interacting} clocks and systems.
\newblock {\em arXiv:1712.00081} {\bf 2017},
\newblock arXiv:1712.00081.

\bibitem{Ber84}
{Berry, M.V.}
\newblock Quantal phase factors accompanying adiabatic changes.
\newblock {\em Proc. Royal Soc. Lond. Math. Phys. Sci.} {\bf 1984}, {\em 392},~45--57,
\newblock
  doi:10.1098/rspa.1984.0023.

\bibitem{Ber85}
Berry, M.V.
\newblock Classical adiabatic angles and quantal adiabatic phase.
\newblock {\em J. Phys. A Math. Gen.} {\bf 1985}, {\em 18},~15,
\newblock
  doi:10.1088/0305-4470/18/1/012.

\bibitem{Kie94}
Kiefer, C.
\newblock The semiclassical approximation to quantum gravity.
\newblock  In \emph{Canonical {Gravity}: {From} {Classical} to {Quantum}}; Ehlers, J.,
  Friedrich, H., Eds.; Lecture {Notes} in {Physics}; Springer: Berlin/Heidelberg, Germany, 1994; pp. 170--212.

\bibitem{Hal85}
Halliwell, J.J.; Hawking, S.W.
\newblock Origin of structure in the {Universe}.
\newblock {\em Phys. Rev. D} {\bf 1985}, {\em 31},~1777--1791,
\newblock
  doi:10.1103/PhysRevD.31.1777.

\bibitem{Bal90}
Balbinot, R.; Barletta, A.; Venturi, G.
\newblock Matter, quantum gravity, and adiabatic phase.
\newblock {\em Phys. Rev. D} {\bf 1990}, {\em 41},~1848--1854,
\newblock
  doi:10.1103/PhysRevD.41.1848.

\bibitem{Dat93}
Datta, D.
\newblock Semiclassical backreaction and berry’s phase.
\newblock {\em Mod. Phys. Lett. A} {\bf 1993}, {\em 8},~191--196,
\newblock
  doi:10.1142/S0217732393000192.

\bibitem{Kie07}
Kiefer, C.
\newblock {\em Quantum {Gravity}}; Oxford University Press: Oxford, UK, 2007.

\bibitem{Bro89}
Brout, R.; Venturi, G.
\newblock Time in semiclassical gravity.
\newblock {\em Phys. Rev. D} {\bf 1989}, {\em 39},~2436--2439,
\newblock
  doi:10.1103/PhysRevD.39.2436.

\bibitem{Ber96}
Bertoni, C.; Finelli, F.; Venturi, G.
\newblock The {Born}-{Oppenheimer} approach to the matter---Gravity system and
  unitarity.
\newblock {\em Class. Quant. Grav.} {\bf 1996}, {\em 13},~2375--2384,
\newblock
  doi:10.1088/0264-9381/13/9/005.

\bibitem{Ven90}
Venturi, G.
\newblock {Minisuperspace, matter and time}.
\newblock {\em Class. Quant. Grav.} {\bf 1990}, {\em 7},~1075--1087,
\newblock
  doi:10.1088/0264-9381/7/6/014.

\bibitem{Joo03}
Joos, E.; Zeh, H.D.; Kiefer, C.; Giulini, D.J.W.; Kupsch, J.; Stamatescu, I.O.
\newblock {\em Decoherence and the {Appearance} of~a~{Classical} {World} in
  {Quantum} {Theory}}, 2nd ed.; Springer: Berlin/Heidelberg, Germany, 2003.

\end{thebibliography}

\end{document}